\documentstyle[12pt,epsf]{article}
\newcommand{\CP}{Chan Paton\ }
\newcommand{\ra}{\rangle}
\newcommand{\eps}{\epsilon}

\newcommand{\II}{{\cal I}}

\newcommand{\NN}{{\cal N}}

\newcommand{\wt}{\widetilde}

\newcommand{\be}{\begin{equation}}
\newcommand{\ee}{\end{equation}}
\newcommand{\ben}{\begin{eqnarray}\displaystyle}
\newcommand{\een}{\end{eqnarray}}
\newcommand{\refb}[1]{(\ref{#1})}
\newcommand{\p}{\partial}
\newcommand{\sectiono}[1]{\section{#1}\setcounter{equation}{0}}

\begin{document}

{}~ \hfill\vbox{\hbox{hep-th/9812031}\hbox{MRI-PHY/P981171}
}\break

\vskip 3.5cm

\centerline{\large \bf BPS D-branes on Non-supersymmetric Cycles}
\medskip

\vspace*{6.0ex}

\centerline{\large \rm Ashoke Sen
\footnote{E-mail: asen@thwgs.cern.ch, sen@mri.ernet.in}}

\vspace*{1.5ex}

\centerline{\large \it Mehta Research Institute of Mathematics}
 \centerline{\large \it and Mathematical Physics}

\centerline{\large \it  Chhatnag Road, Jhoosi,
Allahabad 211019, INDIA}

\vspace*{4.5ex}

\centerline {\bf Abstract}

In certain regions of the moduli space of K3 and Calabi-Yau
manifolds, D-branes wrapped on non-supersymmetric cycles may
give rise to stable configurations. We show that
in the orbifold limit, some
of these stable configurations can be described by solvable
boundary conformal field theories. 
The world-volume theory of N coincident
branes of this type is described by a non-supersymmetric U(N)
gauge theory.
At the boundary of the region
of stability, there are marginal deformations connecting the
non-supersymmetric brane to a pair of D-branes wrapped on
supersymmetric cycles. We also discuss various relationships
between BPS and non-BPS D-branes of type II string theories.

\vfill \eject

\tableofcontents

\baselineskip=18pt

\sectiono{Introduction and Summary} \label{s1}

Type IIA (IIB) string theory contains $2p$-dimensional
($(2p+1)$-dimensional) extended objects known as Dirichlet
branes (D-branes), which are invariant under half of the space-time
supersymmetry transformations of the theory, are charged
under the gauge fields arising in the Ramond-Ramond (RR) sector
of the theory\cite{POLC}, and are stable. Upon compactification
of the type II string theory on a K3 surface or a Calabi-Yau
three fold, we can get stable, supersymmetric branes in the
compactified theory by wrapping these D-branes on various
supersymmetric cycles\cite{BBS} of the compact manifold. 
These branes can also be shown to be stable, and invariant under
part of the supersymmetry transformations of the theory. 

Generically K3 surfaces and Calabi-Yau 3-folds also contain
topologically non-trivial cycles which are not supersymmetric,
and we can consider configurations where a D-brane is wrapped
around one such cycle. The stability of this configuration is not
guaranteed by supersymmetry. Typically such a
non-supersymmetric cycle is homologically equivalent to a sum of
several supersymmetric cycles. Thus it is in principle possible
for a D-brane wrapped on a non-supersymmetric cycle to decay into
several D-branes, each wrapped on a supersymmetric cycle. Whether
such a decay is energetically favourable depends on the values of
various moduli parametrising the vacuum.
By working in the orbifold limit
of K3 and certain Calabi-Yau space we show that D-branes wrapped
on non-supersymmetric cycles can be stable in certain regions of
the moduli space. Furthermore, it is possible to describe such a
wrapped brane by a solvable boundary conformal field theory.
Beyond the region of stability, the state can decay into a pair
of BPS states obtained by wrapping a D-brane on supersymmetric
cycles. This is signalled by the appearance of a tachyonic mode
on the brane world-volume. At the boundary of the region of
stability, the tachyon represents an exactly marginal deformation
of the boundary conformal field theory, and interpolates between
the non-BPS state and a pair of BPS states.

The starting point in our analysis are the non-BPS D-branes of type
II string theories discussed in
\cite{NB,BERGAB,SPINOR,DPART,KTHEORY}.
In particular type IIA (IIB) string theory contains non-BPS
D-$(2p+1)$ (D-$2p$) branes. These branes are unstable, as
indicated by the presence of a tachyonic mode on the brane
world-volume.  We show that
if we start with type IIA string theory on $T^4$,
take the non-BPS D-string of type IIA string
theory wrapped on a circle of the torus, and mod out the theory by 
a $Z_2$ transformation which changes the sign of all the
coordinates of the torus, then the resulting configuration can be
interpreted as a D-membrane of type IIA string theory, wrapped on
a non-supersymmetric cycle of the K3 orbifold. In certain range
of values of the radii of the compact directions, the spectrum of
open strings on the brane is free from tachyons, and hence the
brane is stable. Outside this range a tachyonic mode develops
indicating the existence of a lower energy configuration with the
same quantum numbers. This result can be generalized to branes of
higher dimensions, and also to the case of Calabi-Yau manifold
obtained by taking the
$Z_2\times Z_2$ orbifold of a six dimensional torus.

The paper is organised as follows. Throughout the paper we work
in the limit of weak string coupling and restrict our analysis to
open string tree level. In section \ref{s2} we analyse
some aspects of type IIA D-string wrapped on a circle. This is a
non-BPS configuration and has tachyonic modes. If the radius of
the circle is $R$, and if $x$ denotes the coordinate along the
circle, then $T(x)$ has an expansion:
\be \label{ei1}
T(x)=\sum_{n=-\infty}^\infty T_n e^{inx/R}\, .
\ee
The $n$th mode has mass$^2$ 
\be \label{ei2}
m_n^2={n^2\over R^2}-{1\over 2}\, ,
\ee
in $\alpha'=1$ unit.
Thus at the critical radius $R=\sqrt 2$, $T_{\pm 1}$ becomes
massless. We show that the combination $(T_1-T_{-1})$ represents
an exactly marginal deformation at the critical radius and study
the effect of switching on the vacuum expectation value of
$(T_1-T_{-1})$. This can be done using bosonization techniques
similar to the ones discussed in \cite{SPINOR}. The final result
is that by switching on an appropriate vev of the tachyon field,
the boundary conformal field theory describing the system can be
reduced to the one describing a $D0-\bar D0$ pair of type IIA
string theory, situated at diametrically opposite points on the
circle. 
\begin{figure}[!ht]
\begin{center}
\leavevmode
\epsfbox{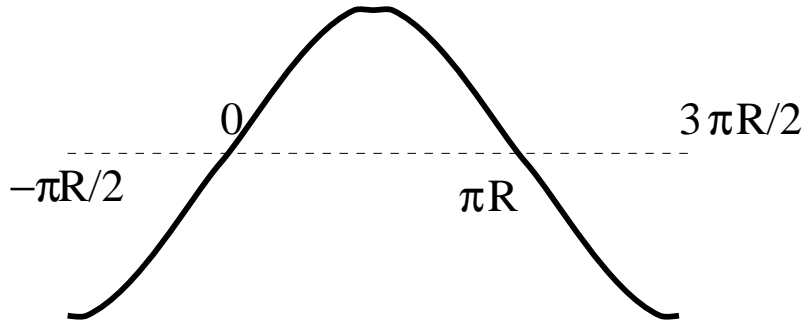}
\caption{The tachyon background corresponding to $(T_{1}-T_{-1})$
excitation.} \label{f1}
\end{center}
\end{figure}

{}From eq.\refb{ei1} we see that switching on $(T_1-T_{-1})$
corresponds to a tachyon field configuration proportional to
$\sin(x/R)$. This has been shown in Fig.\ref{f1}. This diagram
suggests that switching on the tachyon vev of this form corresponds
to the creation of a kink-antikink pair on the D-string.
Combining this with the results stated in the last paragraph, we
see that a kink (antikink) solution on a non-BPS D-string of type
IIA string theory can be identified to a D0-brane (anti-D0-brane)
of type IIA string theory.

If we make an $R\to(1/R)$ duality transformation on the compact
coordinate, then type IIA string theory goes to type IIB
string theory, the initial non-BPS D-string becomes the non-BPS
D-particle, and the final D0-brane anti-D0-brane pair separated
by $\pi R$ becomes a D-string anti-D-string pair of type IIB with
half a unit of Wilson line on one of them. This is precisely the
system studied in \cite{SPINOR}, except that the final state here
was the initial state there, and the initial state here was the
final state there. Thus the results of section \ref{s2} could
have been derived by running the analysis of \cite{SPINOR}
backwards. There is however a specific reason why we have chosen
to carry out the analysis explicitly. Although \cite{SPINOR}
shows how, by starting with a D-string - anti-D-string
configuration of IIB, we can produce a non-BPS D-particle via a
series of marginal deformations, it was never shown explicitly
that this D-particle is identical to the one described in
\cite{DPART}. Thus it is instructive to carry out the analysis
backwards, {\it i.e.} start from a non-BPS D-particle of IIB
as defined in \cite{DPART}, and find the series of marginal
deformations that takes it to a D-string anti-D-string pair. This
is precisely what is established by the analysis of section
\ref{s2}.

In section \ref{s3} we take the non-BPS D-string of type IIA
compactified on a circle, and mod out the theory by a $Z_2$
transformation $\II_4$ which reverses the direction of the circle
(which we denote by $x^9$) and also of three other directions
$x^6,x^7,x^8$. This gives a non-BPS state in the orbifold theory.
We show that the zero momentum mode of the tachyon is projected
out, and so from eq.\refb{ei2} we see that the spectrum is free
from tachyons in the range $R\le \sqrt 2$. Thus this gives a
stable non-BPS state in the theory\footnote{This configuration is
related by T-duality to the stable non-BPS state discussed in
refs.\cite{NB,BERGAB}.}. In order to find a physical
interpretation of this state, we note that $T_1-T_{-1}$ survives
the projection by $\II_4$, and hence  at the critical radius
$R=\sqrt 2$ there still exists a marginal deformation which
takes the present state to a D0-brane $-$ anti-D0-brane pair 
situated
at the fixed points (planes) of the $Z_2$ transformation $\II_4$.
These in turn can be interpreted as type IIA membranes wrapped on
the 2-cycles associated with the blow up modes of the fixed
points\cite{DOUGMOOR}. Since the stable non-BPS state below the
critical radius has the same quantum numbers as the sum of the
quantum numbers of this pair of wrapped membranes, it is natural
to interprete the non-BPS state as a membrane wrapped on a
non-supersymmetric 2-cycle which is homologically equivalent to
the sum of the two 2-cycles associated with the two blow up
modes. As the radius changes from below $\sqrt 2$ to above $\sqrt
2$, it becomes energetically favourable for the brane wrapped on
the non-supersymmetric cycle to break up into a pair of branes
wrapped on supersymmetric cycles. 

If we take the directions $x^6,\ldots x^8$ also to be compact,
then the theory under study is type IIA string theory on a K3
orbifold, and the non-BPS state describes a membrane wrapped on a
non-supersymmetric cycle of K3. This construction can be easily
generalized to a D-$(2p+2)$-brane (D-$(2p+1)$-brane) 
of type IIA (IIB) string
theory wrapped on a non-supersymmetric 2-cycle of K3 by starting
with an initial configuration where we have a non-BPS D-$(2p+1)$
(D-$2p$) brane of IIA (IIB) with one direction along $x^9$ and
other directions along the non-compact directions.

In section \ref{s3a} we study the world-volume theory of $N$
coincident branes of this type. This requires computing the
spectrum of massless open string states with ends on the
D-branes, and can be done by standard
techniques. Thus, for example, if we consider the D5-brane of
type IIB wrapped on a non-supersymmetric 2-cycle of K3,
we get a non-supersymmetric U(N) gauge theory in (3+1) dimensions
with four massless Majorana fermions and two
massless scalars in the adjoint
representation of the gauge group,

In section \ref{s4} we consider some generalisations of the
results of section \ref{s3}. In particular we consider the case
where we have the same K3 orbifold, but the starting
configuration is a non-BPS D3-brane with all three directions
tangential to the torus. Using $R\to(1/R)$ duality
transformations we argue that after modding out the theory by
$\II_4$ this configuration represents a membrane wrapped on a new
2-cycle of K3. We also construct non-BPS states on a Calabi-Yau
manifold obtained by compactifying two more directions ($x^4$ and
$x^5$), and modding out the theory further by a new $Z_2$
transformation $\II_4'$ which reverses the direction of
$x^4,x^5,x^6$ and $x^7$. We include appropriate shifts in the
definition of $\II_4'$ so that neither $\II_4'$ nor $\II_4\II_4'$ 
has any fixed point. (This model was discussed in \cite{FHSV}.) By
starting with an appropriate $\II_4$ and $\II_4'$ invariant
combination of non-BPS D-branes wrapped on various directions of
$T^6$, we construct branes wrapped on non-supersymmetric 2- and
3-cycles of the Calabi-Yau manifold.
\begin{figure}[!ht]
\begin{center}
\leavevmode
\epsfbox{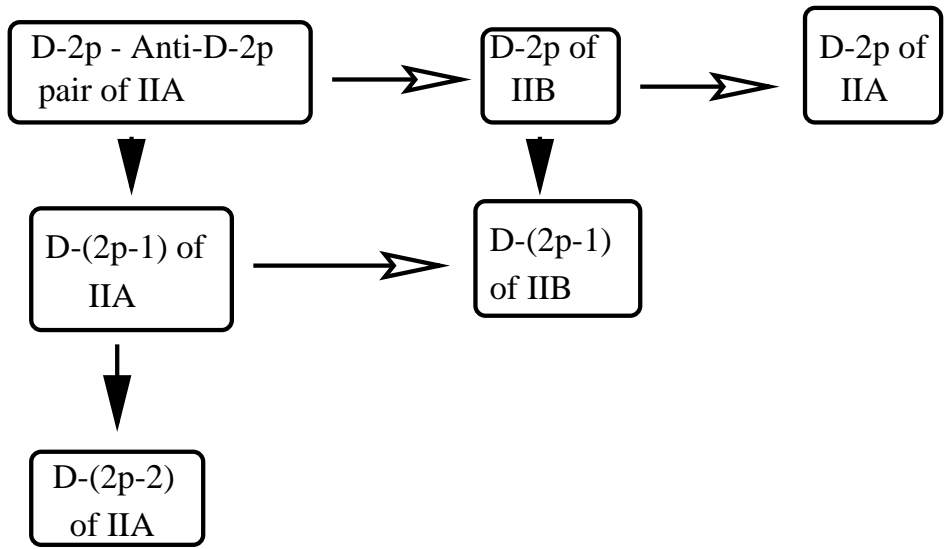}
\caption{The relationship between different D-branes. The
horizontal arrows represent the result of modding out the theory
by $(-1)^{F_L}$, where the vertical arrows represent the effect
of constructing a tachyonic kink solution.
} \label{f2}
\end{center}
\end{figure}

Section \ref{s5} is somewhat outside the main theme of this
paper. In this section we discuss the interrelation between
various supersymmetric and non-supersymmetric D-branes of type II
string theories. These relations take the form of two step  descent
relations where we start from a brane-antibrane pair and end up
with a single BPS D-brane. The first set of relations is obtained as
follows. Let us consider a D-$2p$ $-$ anti-D-$2p$ brane pair of type
IIA string theory. There is a tachyonic excitation on this 
system\cite{TACHYONIC}
such that the tachyonic ground state corresponds to vacuum
configuration\cite{TACH,SRED,GUTPER}. If instead we take a 
tachyonic kink solution on this brane antibrane pair, 
then the analysis of \cite{SPINOR} shows that it describes a
non-BPS D-$(2p-1)$ brane of IIA. This system also has a tachyonic
excitation. If we consider a kink solution on the D-$(2p-1)$ brane
associated with this tachyon, then according to the analysis of
section \ref{s2}, the result is a BPS D-$(2p-2)$-brane of IIA. Similar
result holds if we start with a D-$(2p+1)$-brane $-$
anti-D-$(2p+1)$
brane pair of type IIB string theory. In this case the end
product is a D-$(2p-1)$ brane of IIB.

Another set of relations which we derive in this section
is as follows. Let us again start from a D-$2p$-brane $-$
anti-D-$2p$-brane pair of IIA. But this time, instead of considering
the tachyonic kink solution on this pair, we mod out the theory
by $(-1)^{F_L}$, 
where $(-1)^{F_L}$ acts as $-1$ on all
the Ramond sector states on the left-moving part of the
world-sheet of the fundamental string,
and leaves the other sectors unchanged. 
We show that the result is a non-BPS
D-$2p$-brane of IIB. Upon further modding out the theory by
$(-1)^{F_L}$ we get a BPS D-$2p$-brane of IIA. 

Thus we have two sets of descent relations relating the various
BPS and non-BPS D-branes. These relations have been summarized in
Fig.\ref{f2}.

\sectiono{Type IIA D-string on a Circle} \label{s2}

In this section we shall start with a non-BPS D-string of type
IIA string theory wrapped on a circle, and identify a series of
marginal deformations which map it to a D0-brane anti- D0-brane
pair, situated at diametrically opposite points on the circle. If
we make a T-duality transformation along the circle, then the
starting configuration is a D-particle of type IIB string theory,
and the final configuration is a D-string anti-D-string pair,
with half a unit of Wilson line along one of the strings. It was
already shown in \cite{SPINOR} that these two configurations are
related by marginal deformation; so one might wonder why we
cannot simply take the result of that paper. To this end, note
that in the analysis of \cite{SPINOR}, we started from the
D-string anti-D-string pair of IIB, and identified a series of
marginal deformations whose end product was a D-particle on a
circle. Although there is a great deal of evidence that this
D-particle is the same as the D-particle defined in
refs.\cite{BERGAB,DPART}, this was never proved conclusively. The
analysis of this section starts from the T-dual of the D-particle
described in \cite{BERGAB,DPART} and identifies that marginal
deformation that takes it to the T-dual of the D-string 
anti-D-string pair of IIB. Thus the result of this section can be
taken to be further evidence for the equivalence between the
D-particle of \cite{BERGAB,DPART} and of \cite{SPINOR}.

The non-supersymmetric D-string of type IIA string theory can be
described in a manner similar to the one used in describing the
non-supersymmetric D-particle of type IIB string
theory\cite{DPART}.
If we take the D-string to lie along the 9th
direction, then we put Dirichlet boundary condition on the
coordinates $X^1,\ldots X^8$, and Neumann boundary condition on
the coordinates $X^0, X^9$. However, unlike an ordinary D-string
of type IIB string theory, the spectrum of open strings with both
ends on the type IIA D-string contains Fock space states which are
$(-1)^F$ even as well as $(-1)^F$ odd; with the $(-1)^F$ even
states carrying a Chan Paton (CP) factor equal to the $2\times 2$
identity matrix $I$, and the $(-1)^F$ odd states carrying \CP factor
$\sigma_2$, where $\sigma_i$ ($1\le i\le 3$) are the three Pauli
matrices\footnote{A slightly different description, which is
extremely useful for discussing spinor states, has been given in
\cite{KTHEORY}.
But we shall not use this
description here.}. 
We use the standard convention that $(-1)^F$ acts as
$-1$ on the Neveu-Schwarz (NS) sector ground state of the open
string. This system clearly has a tachyonic excitation coming
from the NS sector ground state in the \CP sector $\sigma_2$.

We shall be analysing the case where the 9th direction has been
compactified on a circle of radius $R_9$ so that the momentum
$k_9$ along this direction is quantized in units of $1/R_9$. 
For simplicity of
notation we shall denote $X^9$, $R_9$ and $k_9$ by $X$, $R$ and
$k$ respectively. In the $\alpha'=1$ unit, the tachyon has
mass$^2=(-1/2)$. Thus the $n$th mode $T_n$
of the tachyon, defined through the expansion:
\be \label{emn1}
T(x) = \sum_{n=-\infty}^{\infty} T_n e^{i{n\over R} x}\, ,
\ee
has effective mass$^2$:
\be \label{e2}
m_n^2={n^2\over R^2}-{1\over 2}\, .
\ee
{}From this we see that at the critical radius 
\be \label{e3}
R={\sqrt 2}
\ee
$T_{\pm 1}$ becomes massless. The vertex operators for $T_{\pm
1}$ are given as follows.
Let $\psi$, $\wt\psi$ denote the right- and the
left-moving components of the world-sheet fermion associated with
the 9th direction, $X_L$, $X_R$ denote the left- and the
right-moving components of the scalar field $X(=X_L+X_R)$, and
$\wt\Phi,\Phi$ denote the left- and the right-moving components
of the bosonized ghost\cite{FMS}. Then,
for NS sector states, the fields $X$, $\Phi$, $\wt\Phi$,
$\psi$, $\wt\psi$
satisfy the boundary conditions:\footnote{These
boundary conditions are written 
in the coordinate system where the open string
world sheet is represented as the upper half plane.}
\be \label{e1}
(X_L)_B=(X_R)_B\equiv X_B/2, \quad \psi_B=\wt\psi_B, 
\quad \Phi_B=\wt\Phi_B \, ,
\ee
where the subscript $B$ denotes the values of various fields
at the boundary of the world-sheet. 
In the $-1$ picture\cite{FMS},
the vertex operator of
$T_{\pm 1}$ at the critical radius is given by:
\be \label{e4}
V^{(-1)}_{\pm} = - e^{-\Phi_B} e^{\pm{i\over \sqrt 2}
X_B}\otimes \sigma_2\, .
\ee
The overall $-$ sign in the above equation is a matter of
convention.
In the
zero picture these vertex operators take the form:
\be \label{e5}
V_{\pm}^{(0)} = \mp i \psi_B e^{\pm{i\over\sqrt 2} X_B}\otimes
\sigma_2\, .
\ee

As in \cite{SPINOR}, we shall now fermionize the world-sheet
scalar $X$ in order to express the vertex operators in a simpler
form.   The bose-fermi relation takes the form:
\be \label{e16}
e^{i\sqrt 2 X_R} = {1\over \sqrt 2} (\xi + i\eta)\, , \qquad
e^{i\sqrt 2 X_L} = {1\over \sqrt 2} (\wt\xi + i\wt\eta)\, .
\ee
We can find another representation of the same conformal field theory by
rebosonizing the fermions as follows:
\be \label{e22}
{1\over \sqrt 2} (\xi + i\psi) = e^{{i \sqrt 2} \phi_R}\, , \qquad
{1\over \sqrt 2} (\wt\xi + i\wt\psi) = e^{{i \sqrt 2} \phi_L}\, .
\ee
$\phi$ represents a free bosonic field with radius $\sqrt 2$.  There is
a third representation in which we use a slightly different 
rebosonization:
\be \label{e22a}
{1\over \sqrt 2} (\eta + i\psi) = e^{{i \sqrt 2} \phi'_R}\, , \qquad
{1\over \sqrt 2} (\wt\eta + i\wt\psi) = e^{{i \sqrt 2} \phi'_L}\, ,
\ee
where $\phi'$ is another scalar field of radius $\sqrt 2$. 
We can also relate the fermionic and the bosonic U(1) currents as
follows:
\be \label{eyy3}
\psi\xi = i\sqrt 2 \p \phi_R\, , \qquad \eta\xi = i\sqrt 2 \p X_R\, ,
\qquad \psi\eta = i\sqrt 2 \p \phi'_R\, .
\ee
There are also
similar relations involving the left-moving currents.

Since $X$ satisfies Neumann boundary 
condition $(X_L)_B=(X_R)_B$
at the boundary of the world
sheet, this translates to the Neumann boundary condition on the
fermions:
\be \label{e17}
\xi_B=\wt\xi_B\, , \qquad \eta_B=\wt\eta_B\, .
\ee
{}From \refb{e1}, \refb{e22}, \refb{e22a} and \refb{e17} we see that
$\phi$ and $\phi'$ both satisfy
Neumann boundary condition at both ends: 
\be \label{exp1}
(\phi_R)_B=(\phi_L)_B\equiv {1\over 2}\phi_B, \qquad
(\phi'_R)_B=(\phi'_L)_B\equiv {1\over 2}\phi'_B\, .
\ee
However, as argued in
\cite{SPINOR}, in the Ramond sector $\phi$,
$\phi'$ satisfy Neumann boundary condition at one end and
Dirichlet boundary condition at the other end.

Using the bosonization relations \refb{e16}-\refb{eyy3} we can
now express the tachyon vertex operators in a simpler form.
However, before we do that, we need to address a subtle issue
related to bosonization. Analysing  \refb{e16} carefully we see
that the left-hand side of this equation commutes with an
operator carrying odd world-sheet fermion number ({\it e.g.}
$\psi$),
whereas the right hand side,
being a fermionic operator, anti-commutes with an operator of odd
world-sheet fermion number. In order to resolve this problem, we
need to assign `cocycle' factors\cite{COCYCLE} 
as follows. 
To every
operator which is odd under $(-1)^F$ we assign a cocycle factor
of $\tau_3$, and to every operator which carries odd unit of
momentum along $x^9$, we assign a cocycle factor of $\tau_1$.
Here $\tau_i$ ($1\le i\le 3$) are Pauli matrices. 
Thus eq.\refb{e16} now takes the form:
\be \label{e16kk}
e^{i\sqrt 2 X_R} = {1\over \sqrt 2} (\xi + i\eta)\otimes \tau_1\, , \qquad
e^{i\sqrt 2 X_L} = {1\over \sqrt 2} (\wt\xi + i\wt\eta)\otimes
\tau_1\, .
\ee
This gives an extra $-$ sign in the commutation relation  of the
right hand side of this equation with the $(-1)^F$ odd states.
Thus both sides of the equation now has the same commutation
relations.\footnote{One might wonder why similar cocycle factors
were not necessary in the analysis of \cite{SPINOR}. There all
allowed vertex operators on the D-string anti- D-string system
were even under $(-1)^Fh$, where $h=-1$ ($+1)$ for vertex
operators carrying odd (even) units of $x^9$ momentum. The
cocycle factors for a $(-1)^F$ even, $h$ even state is $I$,
whereas that for a $(-1)^F$ odd, $h$ odd state is $\tau_2$. Since
these commute with each other, we could ignore them in the
analysis of \cite{SPINOR}.}

Combining this rule with the rule for the \CP factor, 
we see that if $n$ denotes the number of units of
$x^9$ momentum carried by a vertex operator, then we need to
assign the following \CP and cocycle factors depending on the
values of $(-1)^n$ and $(-1)^F$:
\vbox{
\begin{center}
Table 1 \\
\medskip
\begin{tabular}{|c|c|c|}
\hline
$(-1)^F$ & $(-1)^n$ & CP$\otimes$cocycle factor \\
\hline
even & even & $I\otimes I \equiv I$ \\
\hline
even & odd & $I\otimes \tau_1 \equiv \Sigma_3$ \\
\hline
odd & even & $\sigma_2 \otimes \tau_3 \equiv \Sigma_2$ \\
\hline
odd & odd & $\sigma_2 \otimes \tau_2 \equiv \Sigma_1$ \\
\hline
\end{tabular}
\end{center}
}
Note that the $4\times 4$ matrices $I$ and $\Sigma_i$ defined
above satisfy the same algebra as the $2\times 2$ identity matrix
and the Pauli matrices. With the help of these bosonization
rules, and eqs.\refb{e4}, \refb{e5}, we can express
\be \label{ef1}
V_T\equiv {1\over \sqrt 2} (V_+-V_-)\, ,
\ee
representing the vertex operator correponding to the tachyonic
mode $(T_{1} - T_{-1})$, as:
\be \label{ef2}
V_T^{(-1)} = e^{-\Phi_B} \eta_B \otimes \Sigma_1, \qquad
V_T^{(0)} = \psi_B\xi_B \otimes \Sigma_1\, .
\ee
We shall concentrate our attention on the NS sector states, for
which $\phi$ satisfies Neumann boundary condition. In this case,
using eq.\refb{eyy3} and the definition of $\phi_B$ given in
eq.\refb{exp1} we can rewrite
$V_T^{(0)}$ as:
\be \label{ef3}
V_T^{(0)} = {i\over \sqrt 2} \p\phi_B \otimes \Sigma_1\, .
\ee
{}From here on, the analysis proceeds exactly as in
\cite{SPINOR}. Switching on vev of the tachyon amounts to
switching on a Wilson line along the $\phi$ direction. If as in
\cite{SPINOR} we label the tachyon vev by a parameter $\alpha$
with appropriate normalization, 
then at $\alpha=1$ the spectrum takes the
following form.  In the sectors with \CP$\otimes\ $cocycle
factor $I$ or $\Sigma_1$, the spectrum remains unchanged from its
form at $\alpha=0$, whereas
in the sector with \CP$\otimes\ $cocycle factor $\Sigma_2$ and
$\Sigma_3$, the allowed Fock space states have opposite $(-1)^F(-1)^n$
quantum numbers compared to the spectrum at $\alpha=0$. Since
from table 1 we see that in
these sectors, $(-1)^F(-1)^n=-1$ in the absence of tachyon vev,
we conclude that at $\alpha=1$ these states have
$(-1)^F(-1)^n=1$. On the other hand, in the sectors corresponding
to $I$ and $\Sigma_1$, $(-1)^F(-1)^n=1$ in the absence of tachyon
vev, and hence they remain so even at $\alpha=1$. 
Thus we can
conclude that at $\alpha=1$, all states have $(-1)^F(-1)^n=1$;
and furthermore, every state in the Fock space with
$(-1)^F(-1)^{n}=1$ appears twice in the spectrum.
This in turn implies that there is no tachyonic state in the
spectrum at this value of $\alpha$, as the possible tachyon
state, coming from the NS sector ground state, has $(-1)^F=-1$
and $n=0$, and hence has $(-1)^F(-1)^n=-1$.

Following \cite{SPINOR} we can now study the effect of increasing
the radius $R$ beyond its critical value $\sqrt 2$. 
As in \cite{SPINOR} one finds
that for $R>\sqrt 2$
the $\alpha=1$ point represents a local minimum of the
tachyon potential, and the effect of increasing
the radius of the $x^9$ direction to $R=L\sqrt 2$ at $\alpha=1$
can be related to
decreasing the radius of the $\phi'$ coordinate to $\sqrt
2/L=2/R$ in the absence of tachyon vev. Denoting
by $\phi'_D$ the coordinate dual to $\phi'$, one finds that the
$\phi'_D$ coordinate has radius $R/2$; and there is Dirichlet
boundary condition along the $\phi'_D$ direction. Since the
$\phi'_D$ radius goes to $\infty$ as $R\to\infty$, it is
natural to interprete $\phi'_D$ as the 9th direction. Its
associated world-sheet fermion is $\xi,\wt\xi$. Denoting by
$F_{new}$ the new world-sheet fermion number under which
$\xi,\wt\xi$ are odd, and $\psi,\wt\psi,\eta,\wt\eta$ are even,
and by $n_\phi'$ the number of units of $\phi'$ momentum or
equivalently $\phi'_D$ winding, we have the relation:
\be \label{ef4}
(-1)^F (-1)^n = (-1)^{F_{new}} (-1)^{n_\phi'}\, .
\ee
This can be easily verified by studying the action of both sides
on various fields. Thus we see that the spectrum of open string
states contains Fock space states with $(-1)^{F_{new}}
(-1)^{n_\phi'}=1$, with each state in the Fock space satisfying
this relation appearing twice in the spectrum. Note that since
$\phi'_D$ has radius $R/2$, $n_\phi'$ unit of winding charge
corresponds to a total winding charge of $\pi R n_\phi'$.

Let us now compare this with the spectrum of open strings in
type IIA string theory compactified on a circle of radius $R$,
with a D0-brane sitting at $x^9=a$, and an anti-D0-brane sitting
at $x^9=\pi R+a$ where $a$ is some constant. 
In this case the Fock space will contain two
copies of $(-1)^F$ even states from open string with both ends on
the D0-brane or both ends on the anti-D0-brane. Also these states
will carry a total winding charge which is integral multiple of
$2\pi R$, {\it i.e.} even multiple of $\pi R$. On the other hand 
open strings with one end on the D0-brane and the other end on
the anti-D0-brane correspond to Fock space states with
$(-1)^F=-1$, carry winding charge which is odd multiple of $\pi
R$, and also come in pairs due to two different orientations of
the string. If we denote by $\pi R n_\phi'$ the total winding
charge, we see that the full spectrum consists of two copies of
the Fock space states for which $(-1)^F(-1)^{n_\phi'}=1$. This is
exactly identical to the spectrum obtained in the previous
paragraph. Thus we
conclude that the marginal deformation of the wrapped D-string of
the type IIA string theory that we have discussed in this section
takes us to a D-particle anti- D-particle pair of type IIA string
theory situated at diametrically opposite points of the circle.

This result can be interpreted in the following way. Switching on
the real component of the tachyon field corresponding to
$(T_1-T_{-1})$ corresponds to a background tachyon field of the
form:
\be \label{ef5}
T(x) \propto \sin{x\over R}\, .
\ee
As suggested by
Fig.\ref{f1}, this corrsponds to the creation of a
kink-antikink pair separated by $\pi R$. 
On the other hand our analysis shows that
switching on $T_{1}-T_{-1}$ deformation takes us to a
D0-$\bar{\rm D}0$
brane pair, separated by $\pi R$. This suggests that we should
identify the kink
(anti-kink) solution on the type IIA D-string as the type IIA
D-particle (anti- D-particle).

This result has the following consequence. By T-dualizing the
analysis of ref.\cite{SPINOR} one can easily see that the type
IIA D-string can be regarded as a tachyonic kink on the
membrane-antimembrane pair. The excitations on this kink has a
new tachyonic mode, due to the fact that
the tachyon field on the membrane
anti-membrane pair is complex, and hence the kink solution is
unstable. Now we see that the kink solution associated with this
new tachyonic mode is stable, and represents the D-particle of
type IIA string theory. On the other hand a simple topological
analysis shows that the double kink, representing the tachyonic
kink on the kink solution on the membrane anti-membrane pair, is
nothing but the tachyonic vortex on the membrane-antimembrane
pair discussed in \cite{SPINOR,KTHEORY}. 
This establishes the claim that the tachyonic vortex on the
membrane antimembrane pair of type IIA string theory represents a
D-particle. This can be easiliy generalized to show the
equivalence of the vortex solution on a D$(p+2)$-brane anti-
D$(p+2)$-brane to a D$p$ brane $-$ a result which has been
suggested in ref.\cite{NB} and used recently to show that D-brane
charges take values on the K-theory of space-time\cite{KTHEORY}.

\sectiono{Type IIA D-string on an Orbifold} \label{s3}

In this section we shall analyse a $Z_2$ orbifold of the system
considered in section \ref{s2}. The $Z_2$ group will be
generated by an element $\II_4$ which reverses the sign of
the 9th direction as well as three other directions which we
shall take to be along $x^6,x^7$ and $x^8$. Also for
definiteness, we shall take the fixed planes of this
transformation to be located at $x^6=x^7=x^8=0$,
$x^9=0,\pi R$.  In order to see the
fate of the non-supersymmetric D-string along the 9th direction
under this orbifold operation, we need to know how $\II_4$ acts
on the open string states with ends lying on the D-string. This,
in turn, requires knowing the action of $\II_4$ on the
oscillators, the momenta, as well as on the vacuum. The
action on the oscillators and the momentum components
is straightforward; those associated
with $X^6,\ldots X^9$ and their world-sheet fermionic partners
change sign; whereas those associated with $X^0,\ldots
X^5$ and their fermionic partners do not change sign. The action
on the vacuum needs to be determined separately in each
\CP sector. First of all in the \CP sector $I$, the
mode corresponding to translation of the D-string along the
$X^i$ direction for $1\le i\le 5$
must be even under $\II_4$ since it has a non-vanishing 2-point
function with the $0i$ component of the space-time metric which
is even under $\II_4$. Since the
$X^1,\ldots X^5$ oscillators do not change sign under $\II_4$,
this shows that $\II_4$ must leave the vacuum in this sector
invariant. In order to determine the action of $\II_4$ on the
vacuum in the \CP sector $\sigma_2$, let us note that if $A_\mu$
denotes the RR sector vector field of the type IIA string theory,
then the two point function of $A_9$, and the tachyonic open
string state in the \CP sector $\sigma_2$, carrying zero momentum
along $x^6,\ldots x^9$, is
non-vanishing. (This can be seen by a computation very similar to
one done in ref.\cite{KTHEORY} for non-supersymmetric D0-brane.)
Since $A_9$ is odd under $\II_4$, this shows that the tachyonic
ground state with zero $x^6,\ldots x^9$ 
momentum is odd under $\II_4$. In other words, the vacuum in the
\CP sector $\sigma_2$ is odd under $\II_4$.

{}From this we see that the tachyon carrying zero momentum along
$x^9$ is projected out when we mod out the theory by $\II_4$. 
For non-zero $x^9$ momentum
$n/R$, the modes which survive are:
\be \label{ef6}
T_n - T_{-n}\, .
\ee
{}Since the $n=0$ mode is absent,
from eq.\refb{e2} we now see that below the critical radius
$R=\sqrt 2$, there are no tachyonic modes. Thus the configuration
is stable in this region. On the other hand above the critical
radius, this configuration becomes unstable due to the appearance
of a tachyonic mode $T_{1}-T_{-1}$. As we have seen in
section \ref{s2}, this signals the possibility of
the decay of the wrapped D-string
into a pair of $D0$-branes at diametrically opposite
points on the $x^9$ axis. In order to see what this final
configuration corresponds to in the orbifold theory, 
we need to have more precise information
about the
locations of the D0-brane - anti-D0-brane pair. Since modding out
by $\II_4$ breaks translation invariance along $x^9$, it is no
longer sufficient to just say that they are separated by a
distance $\pi R$. To this end, note that since we have a single
D0-brane and a single anti- D0-brane, the only way to get an
$\II_4$ invariant configuration is to have one of them located at
$x^9=0$, and the other one located at $x^9=\pi R$. This can also
be seen explicitly from Fig.\ref{f1}. The location of the
0-branes 
after tachyon condensation can be taken to be the places where the
tachyon field vanishes, since these are the places where the
fundamental property of the D-brane $-$ that open strings can end
there $-$ remains intact. From Fig.\ref{f1} we see that when the
combination $T_1-T_{-1}$ is switched on, these
points are located at $x^9=0$ and $x^9=\pi R$. Thus these are the
locations of the D0-brane and the anti-D0-brane after tachyon
condensation.

The effect of modding out such a configuration by $\II_4$ was
studied in \cite{DOUGMOOR}. The result was that these correspond
to membranes wrapped on the collapsed 2-cycles associated with
the orbifold singularities. Naively one would think that
these configurations would have vanishing mass; however this does
not happen since on each of these collapsed 2-cycles there is
half a unit of flux of the rank two anti-symmetric tensor field
arising in the NSNS sector of the closed string\cite{ASPINFLUX}.
Thus we see that when $R>\sqrt 2$, our original configuration
becomes unstable against decay into a pair of membranes, wrapped
around the pair of 2-cycles associated with the orbifold fixed
planes at $(x^5,\ldots x^9)=(0,0,0,0)$ and $(0,0,0,\pi R)$. This
clearly suggests that the stable configuration for $R<\sqrt 2$
that we had started with has the interpretation of a membrane
that wraps around a 2-cycle which is homologically identical to
the sum of the 2-cycles associated with the two fixed points.
This is not a supersymmetric cycle\cite{BBS}, 
as the state obtained by
wrapping a membrane on it is not a BPS state. Nevertheless a
membrane wrapped on it is stable in some region of the moduli
space.

This interpretation can also be seen by constructing the boundary
state\cite{BOUNDARY,BOUNT} corresponding to the original 
configuration. Before the
orbifold projection the boundary state is given by:
\be \label{ef7}
|\theta,U\rangle_{NSNS}\, ,
\ee
where $|\theta,U\ra_{NSNS}$ has been defined in eq.(3.4) of
\cite{NB}. Here $\theta$ is the Wilson line associated with the
U(1) gauge field living on the D-string. Note that although the
boundary states for D-string were defined in \cite{NB} for type
IIB D-strings, we can use the NSNS sector components of the same
boundary state to describe the D-string of type IIA. Comparing
with eq.(3.6), (3.7) of \cite{NB} we see that in the present case
there is no RR component in the boundary state, whereas the NSNS
sector contribution has an extra factor of $\sqrt 2$. This is
consistent with the prescription of \cite{DPART}. When we mod out
by the transformation $\II_4$, $\theta$ is constrained to take
value 0 or $\pi$, and we also need to add extra terms to the
boundary state which will be responsible for projecting out the
$\II_4$ non-invariant states from the open string sector. These
extra terms can also be described in the language of
ref.\cite{NB}. The resulting boundary state describing the type
IIA D-string on the orbifold is given by:
\be \label{ef8}
|\theta,\eps\ra = {1\over \sqrt 2}|\theta,U\ra_{NSNS} 
+ {1\over 2} \eps (|T_1\ra_{RR} + e^{i\theta} |T_2\ra_{RR})\, .
\ee
where $\eps$ takes values $\pm1$ and $|T_1\ra_{RR}$,
$|T_2\ra_{RR}$ are the RR
components of the twisted sector boundary states located at
$x^9=0$ and $x^9=\pi$ respectively, as defined in
\cite{NB}. Note that although in \cite{NB} we considered the case
of type IIB string theory instead of type IIA, and considered
orbifolding by $(-1)^{F_L}\II_4$ instead of just $\II_4$, these
two differences compensate each other so that the states
$|T_i\ra_{RR}$ are valid states ({\it i.e.} satisfy the GSO
projection) in the type IIA orbifold that we
are analysing here.\footnote{This can be seen by noting that the
GSO projection operators in the closed string sector of the 
present theory, acting on the twisted sector RR ground state,
takes the same form as given in eq.(2.41) of \cite{NB}. The
T-dual version of this result has been discussed in
\cite{BERGAB}.}
Following the analysis of \cite{NB} one can easily verify that 
the
effect of adding the twisted sector contribution to the boundary
state is to add an extra projection operator $(1 + \II_4 \cdot
(-1)^F)/2$ to
the open string partition function with $\II_4$ acting as
$(+1)$  and $(-1)^F$ acting as $(-1)$
on the vacuum of the NS sector.
Since
for \CP factor $I$ the allowed states are $(-1)^F$ even, they must
also be $\II_4$ even. On the other hand in the sector with \CP
factor $\sigma_2$ the allowed Fock space states are $(-1)^F$ odd
and hence they must also be $\II_4$ odd. This is equivalent
to assigning
the vacuum $\II_4$ charge $(-1)$ and keeping $\II_4$ even
states in the spectrum. Thus the spectrum of open strings
associated with the boundary state \refb{ef8} agrees with the
spectrum on the non-BPS string described earlier.

{}From \refb{ef8} we see that the type IIA D-string on this
orbifold acts as sources of the RR gauge fields in the twisted
sector. Since these gauge fields couple to membranes wrapped on
the collapsed 2-cycles, we see that the D-string on the
orbifold can be regarded as membrane wrapped simultaneously 
on these two 2-cycles. Since both $\eps$ and $e^{i\theta}$ can
take values $\pm 1$, we can get four different configurations,
representing four possible ways of wrapping the membrane around
the two 2-cycles. Note however that in each case, the
configuration is neutral under the untwisted sector RR gauge
field $A_\mu$
which couples to the D0-brane charge of type IIA string
theory. 

Clearly the above procedure could be repeated even if we
compactify the directions $x^6,\ldots x^8$ so that the orbifold
describes  a K3 manifold. 
Thus the procedure described here gives
us a solvable boundary conformal field theory describing
membranes wrapped on non-supersymmetric cycles of K3. 
However in this case the Fock space
ground state of the open strings with ends on the D-string and
carrying one unit of winding number along $x^6$, $x^7$ or $x^8$,
is not projected out, as $\II_4$ maps a winding number 1 state to
a winding number $-1$ state. If we consider the state with one
unit of winding number along $x^8$ for definiteness, then the
mass $m$ of this state is given by:
\be \label{ekx1}
m^2=R_8^2 -{1\over 2}\, .
\ee
Thus in order that there are no tachyonic excitations, we need to
have
$R_8\ge {1\over \sqrt 2}$.
Similarly by analysing the winding modes along $x^6$ and $x^7$ we
get similar constraints on $R_6$ and $R_7$.
Thus the membrane wrapped on the non-BPS cycle is stable in the
region:
\be \label{ekx3}
R_6\ge {1\over \sqrt 2}, \quad R_7\ge {1\over \sqrt 2}, \quad
R_8\ge {1\over \sqrt 2}, \quad R_9\le \sqrt 2.
\ee
As we have already seen,
beyond the boundary $R_9=\sqrt 2$, the state can decay into a
pair of membranes wrapped on supersymmetric cycles. In order to
see what happens at $R_8=(1/\sqrt 2)$, we note that this
configuration is related to the one at $R_9=\sqrt 2$ by
a T-duality transformation $R_8\to (1/R_9)$, $R_9\to (1/R_8)$.
Since this T-duality transformation maps a supersymmetric cycle
to a supersymmetric cycle, we see that even beyond the boundary
at $R_8=(1/\sqrt 2)$ the non-BPS state becomes unstable against
decay into a pair of D-branes wrapped on supersymmetric cycles of
K3. The same result holds beyond the boundaries at $R_6=(1/\sqrt
2)$ and $R_7=(1/\sqrt 2)$. Note however that the pair of
supersymmetric cycles involved are different at different
boundaries. 

Instead of starting with a non-BPS string of type IIA string
theory, we could have started with a non-BPS $(2p+1)$-brane of
type IIA string theory with $(p\le 2)$, with one of the
tangential directions along $x^9$, and the other directions along
$x^1,\ldots x^{2p}$.\footnote{A non-BPS $(2p+1)$-brane of type
IIA string theory is defined in the
same way as a non-BPS D-string, except that we impose Neumann
boundary condition along $(2p+1)$ spatial directions instead of
only one spatial direction.} After modding out by $\II_4$ this
describes a stable $2p$-brane along non-compact directions in the
range of parameters given in \refb{ekx3}. Following the same
analysis given in this section, one can easily see that this
describes a BPS $(2p+2)$-brane of type IIA string theory, wrapped
on a non-supersymmetric 2-cycle of K3. Similarly, if we start
with a non-BPS D-$2p$-brane of type IIB string theory stretched
along $x^9,x^1,\ldots x^{2p-1}$ ($p\le 3$) and mod out the theory
by $\II_4$, we shall get a D-$(2p+1)$-brane of type IIB, wrapped
on a non-supersymmetric 2-cycle of K3. The case where more than
one tangential directions of the brane are along the compact
directions $x^6,\ldots x^9$ will be discussed in section
\ref{s4}.

\sectiono{World-volume Theory on Coincident D-branes} \label{s3a}

In this section we shall discuss the world-volume theory of the
D-branes constructed in section \ref{s3}. For definiteness we
shall carry out the analysis for a non-BPS 4-brane of type IIB
stretched along $x^1,\ldots x^3,x^9$. After compactification of
$x^6,\ldots x^9$ directions
and modding out by $\II_4$, this will give rise to a type IIB
D5-brane wrapped on a non-supersymmetric 2-cycle, with a
(3+1) dimensional world-volume field theory. 
The results for other cases
can be easily found from this via dimensional
reduction/oxidation.

We begin with the case of a single D-brane of this kind. The
spectrum of massless fields before projection by $\II_4$ can be
easily found. The spectrum of states from the \CP sector $I$ is
that of an $\NN=4$ supersymmetric U(1) gauge theory, {\it i.e.}
it has besides the gauge fields, four Majorana fermions and six
real scalar fields. The spectrum of states from the \CP sector
$\sigma_2$ does not contain any massless bosonic states; however
the Ramond sector gives another four massless Majorana fermions.
Thus we have a U(1) gauge field, eight massless Majorana fermions
and six scalars. Five of these six scalars correspond to the
freedom of moving the brane along its transverse directions
$x^4,\ldots x^8$, while the sixth scalar denotes the component of
the gauge field on the 4-brane along $x^9$.
Upon modding out the theory by $\II_4$, only two of the six
scalars, corresponding to the freedom of moving the branes along
$x^4$ and $x^5$ directions,
survive. The gauge field as well as four of the
eight Majorana fermions also survive the projection. Thus the
final spectrum contains a gauge field, four Majorana fermions and
two scalar fields.

We can now study the effect of bringing 
together $N$ such branes on
top of each other. If the branes are all identical ({\it i.e.} the
parameters $\epsilon$ and $\theta$ appearing in \refb{ef8} are
identical for all branes) then the spectrum of open strings with
two ends on two different branes is identical to that of open
strings with both ends on the same brane. As a result the
spectrum contains
$N^2$ copies of all the fields that appear on the world volume of
a single brane. They describe a non-supersymmetric U(N) gauge
theory with four Majorana fermions and two scalars in the adjoint
representation of the gauge group.

\sectiono{Some Generalizations} \label{s4}

\subsection{Branes wrapped on other non-supersymmetric cycles of
K3}

In the last section we constructed non-BPS states in IIA on K3 by
starting from a non-BPS string wrapped on a circle of the torus,
and then modding out the theory by $\II_4$. But we could also
have started with a non-BPS three brane of IIA 
wrapped on three
of the circles of $T^4$ and modded out the resulting
configuration by $\II_4$. For definiteness, let us assume that
the three brane is wrapped along the 6-7-8 cycle.
Since this configuration is related to
the configuration discussed in section \ref{s3} by a
T-duality transformation $R_i\to (1/R_i)$ for $6\le i\le 9$,
we do not need to carry out the analysis all over again. First of
all, from \refb{ekx3} we see that this configuration is stable in
the region:
\be \label{ekx4}
R_6\le {\sqrt 2}, \quad R_7\le {\sqrt 2}, \quad
R_8\le {\sqrt 2}, \quad R_9\ge {1\over \sqrt 2}.
\ee
In the analysis of section \ref{s3} we have seen that beyond
the region of stability, the non-BPS state discussed there
decays into a pair of
BPS states obtained by wrapping BPS D-branes on supersymmetric
2-cycles of K3. Since the T-duality transformation discussed here
maps a supersymmetric 2-cycle into a supersymmetric 2-cycle, we can
conclude that even in the present case, the non-BPS state decays
into a pair of BPS D2-branes wrapped on supersymmetric 2-cycles
beyond the region of stability. 
The new supersymmetric cycles are related to
the ones in section \ref{s3}
by the T-duality transformation $R_i\to (1/R_i)$ for $6\le i\le
9$.  Thus the non-BPS
configuration obtained by wrapping a 3-brane on the 6-7-8
direction can be interpreted as a D2-brane wrapped on the
non-supersymmetric cycle which is homologically identical to the
sum of the two supersymmetric 2-cycles into which it decays beyond
the region of stability.

As in section \ref{s3},
instead of starting with a non-BPS 3-brane, we could have started
with a non-BPS $(2p+3)$ brane stretched along
$x^6,x^7,x^8,x^1,\ldots x^{2p}$ ($p\le 2$). After modding out by
$\II_4$ this configuration can be interpreted as a BPS $(2p+2)$
brane of IIA, wrapped on the non-supersymmetric cycle discussed
above.

\subsection{Branes wrapped on non-supersymmetric
2- and 3-cycles of a Calabi-Yau
manifold}

In this section we shall discuss construction of branes wrapped
on non-supersymmetric 2- and 3-cycles of a Calabi-Yau manifold.
The specific Calabi-Yau manifold that we shall be considering is
the one discussed in \cite{FHSV}. We compactify type IIA string
theory on $T^6$ labelled by coordinates $x^4,\ldots x^9$, and mod
out the theory by a $Z_2\times Z_2$ symmetry, generated by
$\II_4$ discussed in section \ref{s3}, together with
\be \label{ekx5}
\II_4': (x^4,\ldots x^9)\to (-x^4,-x^5,-x^6+\pi R_6,-x^7,x^8+\pi
R_8,x^9)\, .
\ee
Here $R_n$ denotes the radius of the circle spanned by $x^n$.
This model can be regarded as the result of modding out by
$\II_4'$ the product of the K3 orbifold discussed in
section \ref{s3}
and the two dimensional torus $T^2$ spanned by $x^4,x^5$.
Since $\II_4'$ includes a shift by $\pi R_8$ along $x^8$, there
are no fixed points on $T^6$ under $\II_4'$. Similarly one can
check that there are no fixed points of the transformation
$\II_4\II_4'$ either. Thus the non-trivial cycles of the
Calabi-Yau orbifold
are obtained by taking $\II_4'$ invariant cycles on the product
of $T^2$ and the K3 orbifold discussed in the last section. In
particular if $C$ denotes a two cycle of K3, $C'$ denotes
its image
under the part of $\II_4'$ that acts on $x^6,\ldots x^9$, 
$S^1$ denotes a 1-cycle on $T^2$ labelled by $x^4,x^5$, and
$S^{1\prime}$ denotes its image under the part of $\II_4'$ acting on
$T^2$, then
$C+C'$ gives a
2-cycle of the Calabi-Yau manifold, and
$(C\times S^1+C'\times S^{1\prime})$ gives a 
3-cycle of the Calabi-Yau manifold.

In order to construct a non-BPS D-brane configuration in this
theory, we must start with a non-BPS D-brane configuration on the
original torus which is invariant under $\II_4$ and $\II_4'$. The
non-BPS D-string lying along $x^9$ at
\be \label{eky1}
x^6=x^7=x^8=0, \qquad x^i = a^i \quad \hbox{for} \quad 1\le i\le
5 \, ,
\ee
is invariant under $\II_4$, but transforms under $\II_4'$ to
another non-BPS D-string along $x^9$, located at,
\be \label{eky2}
x^6=\pi R_6, \quad x^7=0, \quad x^8=\pi R_8, \quad x^i = a^i 
\quad \hbox{for} \quad 1\le i\le 3, \quad x^i=-a^i \quad
\hbox{for} \quad i=4,5 \, .
\ee
Here $a^i$ are arbitrary constants.
Thus if we start with a pair of non-supersymmetric D-strings in
type IIA on $T^6$ located at \refb{eky1}, \refb{eky2},
and mod out the theory by the $Z_2\times Z_2$ symmetry, we get a
non-BPS state of type IIA string theory on a Calabi-Yau orbifold.
The boundary states corresponding to the two D-strings must be
chosen in such a way that they are mapped to each other under the
action of $\II_4'$. This relates the $\epsilon$ and $\theta$
parameters of the two D-strings. Given the boundary states,
the spectrum of open strings with ends on the D-string or its
image under $\II_4'$ can be computed in a straightforward manner.

The physical interpretation of this stable non-BPS state is
straightforward. From our analysis of section \ref{s3} we
know that after modding out by $\II_4$, the D-string lying at
\refb{eky1} denotes a membrane wrapped on a non-supersynmmetric
cycle $C$. Since \refb{eky2} is the image of \refb{eky1} under
$\II_4'$,
the D-string lying at \refb{eky2} denotes a membrane wrapped on
$C'$, where $C'$ is the image of $C$ under $\II_4'$. Thus the
combined system of D-strings denotes a membrane wrapped on
$C+C'$, which, according to our previous analysis, is a 2-cycle
of the Calabi-Yau manifold. Thus the D-string configuration
considered here denotes a membrane wrapped on a
non-supersymmetric 2-cycle of the Calabi-Yau manifold.

Instead of starting with a D-string, we could have started with a
pair of D3-branes along $x^9$, $x^1,x^{2}$, located at 
\be \label{eyy1}
x^6=x^7=x^8=0, \qquad x^i = a^i \quad \hbox{for} \quad 3\le i\le
5 \, ,
\ee
and
\be \label{eyy2}
x^6=\pi R_6, \quad x^7=0, \quad x^8=\pi R_8, \quad x^3 = a^3, 
\quad x^i=-a^i \quad
\hbox{for} \quad i=4,5 \, .
\ee
This describes a 2-brane in the non-compact directions and has
the interpretation of a type IIA 4-brane wrapped along a
non-supersymmetric 2-cycle of the Calabi-Yau manifold.

Instead of taking the pair of three branes to be lying parallel
to the 1-2-9 plane, we could have taken them to be parallel
to the 1-5-9 plane, and located at
\be \label{eyyy3}
x^6=x^7=x^8=0, \qquad x^i = a^i \quad \hbox{for} \quad 2\le i\le
4 \, ,
\ee
and
\be \label{eyy4}
x^6=\pi R_6, \quad x^7=0, \quad x^8=\pi R_8, \quad x^i = a^i
\quad \hbox{for} \quad i=2,3,
\quad x^4=-a^4 \, .
\ee
In this case, before the $\II_4'$ projection, the two
D-branes have the interpretation of a D4-brane of type IIA
wrapped on $C\times S^1$ and $C'\times S^{1\prime}$ respectively,
with $S^1$, $S^{1\prime}$ denoting a one cycle on $T^2$ 
along the $x^5$ direction and
its image under $\II_4'$ respectively. Thus the combined system
denotes a type IIA 4-brane wrapped on $(C\times S^1+C'\times
S^{1\prime})$. From our previous discussion we see that after
modding out the theory by $\II_4'$ this denotes a type IIA
4-brane wrapped on a non-supersymmetric 3-cycle of the Calabi-Yau
manifold.

\sectiono{Descent Relations Among D-branes} \label{s5}

The spectrum of D-branes in type IIA string theory
contains BPS $2p$ branes and non-BPS $(2p+1)$ branes. Type IIB
string theory on the other hand contains BPS $(2p+1)$-branes and
non-BPS $2p$-branes.
Using the results of section \ref{s2}, and of \cite{SPINOR},
we get the following set of relations between
different D-branes in type II string theory. 
For definiteness we shall
focus on type IIA string theory, and start with a coincident
$2p$-brane - anti-$2p$-brane pair.
There is a complex tachyon field living on the world-volume of this
system. If the tachyon condenses to the minimum value $T_0$ 
of the tachyon potential everywhere, then this system is
indistinguishible from vacuum. If instead we consider a tachyonic
`kink' solution that is independent of time, as well as
$(2p-1)$ of the spatial coordinates on the world-volume, and has
the following behaviour as a function of the remaining world-volume
coordinate $x$:
\be \label{ez1}
T(x)\to T_0 \quad \hbox{as}\quad x\to\infty, \qquad
T(x)\to -T_0 \quad \hbox{as}\quad x\to-\infty, 
\ee
then it describes a non-BPS D-$(2p-1)$ brane of type IIA string
theory\cite{SPINOR}. 
This non-BPS brane in turn has a real tachyon field $\wt T$ 
living on it, At the minimum $\wt T_0$ of the tachyon potential
the system is again indistinguishible from the vacuum. However if
we consider a time independent
kink solution associated with this new tachyon
which is independent of $(2p-2)$ of the spatial directions on the
world-volume and behaves as
\be \label{ez2}
\wt T(y)\to \wt T_0 \quad \hbox{as}\quad y\to\infty, \qquad
\wt T(y)\to -\wt T_0 \quad \hbox{as}\quad y\to-\infty, 
\ee
$y$ being the remaining world-volume coordinate, then it can be
identified as the BPS $(2p-2)$-brane of the type IIA string theory.

A similar relationship holds between the BPS and non-BPS branes
of type IIB string theory. Using these relations, all
D-branes of type IIB string theory can be regarded as solitons on
9-brane $-$ anti-9-brane system. 
This fact has been used recently to
show that the D-brane charge takes value on the K-theory of
space-time\cite{KTHEORY}.
Similarly all D-branes of type IIA
string theory can be regarded as solitons on an appropriate
8-brane anti-8-brane system. 

In this section we shall discuss another kind of relationship
between BPS and non-BPS D-branes. Our starting point will again
be a brane-antibrane pair. Again for definiteness we shall
consider a $2p$-brane - anti$-2p$-brane pair of type IIA string
theory. But this time, instead of considering a tachyonic soliton
on this pair, let us consider the effect of modding out the
theory by $(-1)^{F_L}$.
Since $(-1)^{F_L}$
changes the sign of all the RR sector fields, and since BPS
D-branes carry RR charge, a BPS D-brane gets transformed to its
antibrane under $(-1)^{F_L}$. As a result the brane-antibrane
pair is invariant under $(-1)^{F_L}$ and it makes sense to mod
out the configuration by $(-1)^{F_L}$.

In the bulk, modding out type IIA string theory by $(-1)^{F_L}$
gives type IIB string theory. The question we shall be interested
in is: what happens to the brane-antibrane system under this
modding? For this we
need to compute the spectrum of open string states on the
brane-antibrane system after the
$(-1)^{F_L}$ projection. 
Before the projection, the open string states are
labelled by $2\times 2$ Chan Paton factors. If we take the
basis of Chan Paton factors to be the identity matrix $I$ and the
three matrices $\sigma_1$, $i\sigma_2$ and $\sigma_3$,
then, since $I$ and $\sigma_3$
correspond to open strings with both ends lying on the D-brane
(anti-D-brane), they contain Fock space states which are even
under $(-1)^F$. On the other hand, $\sigma_1$ and $\sigma_2$,
representing open string states with one end on the D-brane and
the other end on the anti-D-brane, contain Fock space states
which are odd under $(-1)^F$.
Since in the Neveu-Schwarz-Ramond
formalism $(-1)^{F_L}$ does not act on any of the fields on the
world-sheet of the fundamental string, in
order to determine the spectrum of $(-1)^{F_L}$ invariant open
string states we
only need to determine its action on the Chan-Paton factor
$\Lambda$. 
Since $(-1)^{F_L}$ maps the brane to the antibrane,
its action on the Chan Paton factors takes the form:
\be \label{efz1}
\Lambda\to S\Lambda S^{-1}\, ,
\ee
where $S$ can be taken to be
either $\sigma_1$ or $\sigma_2$. Both choices are
equivalent; for definiteness let us choose $S$ to be $\sigma_2$.
Thus only those open string states for which $\Lambda$ commutes
with $\sigma_2$ survive the projection. This gives:
\be \label{efz2}
\Lambda=I, \sigma_2\, .
\ee
Since the sector $I$ contains $(-1)^F$ even Fock space states,
and the sector $\sigma_2$ contains $(-1)^F$ odd Fock space
states, we see that the spectrum of open strings after the
$(-1)^{F_L}$ projection agrees precisely with the spectrum of a
non-BPS D-$2p$-brane of type IIB string theory. Thus we conclude
that the D-$2p$ anti-D-$2p$-brane pair of type IIA string
theory, after being modded out by $(-1)^{F_L}$, gives a non-BPS
D-$2p$-brane of type IIB string theory. A similar analysis shows
that the D-$(2p+1)$ anti-D-$(2p+1)$-brane pair of type IIB string
theory, after being modded out by $(-1)^{F_L}$, gives a non-BPS
D-$(2p+1)$-brane of type IIA string theory.

Let us now consider the effect of modding out the non-BPS
D-$2p$-brane of type IIB string theory by $(-1)^{F_L}$. This is a
possible operation,
since this $2p$-brane does not carry any RR charge, and hence is
invariant under $(-1)^{F_L}$. In the bulk, the theory goes back to
type IIA string theory. In order to see what happens to the
D-brane, we need to find the spectrum of open strings after the
$(-1)^{F_L}$ projection. As discussed earlier, $(-1)^{F_L}$ does not
act on the fields on the world-sheet of the fundamental string, 
and hence we only need to find its action on the Chan Paton
factors. It turns out that $(-1)^{F_L}$ leaves states in the Chan
Paton sector $I$ invariant, and reverses the sign of the states
in the Chan Paton sector $\sigma_2$. This can be seen by noting
that {\it (i)} the two point function of the anti-symmetric tensor
field $B_{\mu\nu}$ in the NSNS sector of the closed string, and
the gauge field ${\cal A}_\mu$ arising in the Chan Paton
sector $I$ of the open string, is non-vanishing, and {\it (ii) }
the two point function of the $2p$-form field $A^{(2p)}$ 
arising in the RR sector of the closed string, and the tachyonic
open string state arising in the Chan Paton sector $\sigma_2$, is
non-vanishing. Since $B_{\mu\nu}$ is even under $(-1)^{F_L}$, so
must be states in the Chan Paton sector $I$. On the other hand
since $A^{(p)}$ is odd under $(-1)^{F_L}$, so must be the states
in the Chan Paton sector $\sigma_2$. Thus after the $(-1)^{F_L}$
projection we are only left with states in the Chan Paton sector
$I$. This sector contains Fock space states which are even under
$(-1)^F$. This spectrum agrees with the spectrum of open strings
with both ends on a D-$2p$-brane of type IIA
string theory. Thus we conclude that the result of modding out
the type IIB D-$2p$-brane by $(-1)^{F_L}$ is a BPS D-$2p$-brane
of type IIA string theory. A similar analysis shows that the
result of modding out
the type IIA D-$(2p+1)$-brane by $(-1)^{F_L}$ is a BPS 
D-$(2p+1)$-brane of type IIB string theory. 

These results have been summarized in Fig.\ref{f2}.

\end{document}